\documentclass{rdm}

\def\pd{\hspace{4mm}}

\normallatexbib

\renewcommand{\theequation}
{\arabic{section}.\arabic{equation}}

\begin{document}
\thispagestyle{empty}
\vspace*{2.5in}
\setcounter{page}{0}
\begin{center}
{\Large\bf W1 and W2 Theories, and Their Variants: \\
Thermochemistry in the kJ/mol \\Accuracy Range\\}
\vspace*{0.33333in}
{\large Jan M.L. Martin and S. Parthiban}\\
~\\
{\it Department of Organic Chemistry,\\
Weizmann Institute of Science,\\IL-76100 Re\d{h}ovot, Israel\\}
E-mail: {\tt comartin@wicc.weizmann.ac.il}
\end{center}
\vspace*{0.16666in}
\begin{center}
Chapter 2, pp. 31--65, In:\\ {\large\it Quantum Mechanical Prediction of Thermochemical Data},
\\ed. J. Cioslowski and A. Szarecka;\\ {\em Understanding Chemical
Reactivity} Series, Vol. {\bf 22};\\ Kluwer Academic Publishers, Dordrecht, The
Netherlands, 2001;\\ ISBN {\bf 0-7923-7077-5};\\ 
\end{center}

\setcounter{chapter}{1}
\vspace{-3mm}
\articletitle{W1 and W2 Theories, and Their Variants: \\
Thermochemistry in the kJ/mol Accuracy Range}
\chaptitlerunninghead{W1 and W2 theories} 
\vspace{11pt}
\author{Jan M.L. Martin and S. Parthiban}
\affil{Department of Organic Chemistry, Weizmann Institute of Science, Kimmelman\\
\hbox{Building, IL-76100 Re\d{h}ovot, Israel}} 
\vspace{38pt}
\pagenumbering{arabic}
\setcounter{page}{31}
\section{\, Introduction and background}
\vspace{8pt}
\pd The last fifteen years witnessed the development of a number of "black-box" computational thermochemistry
methods. 
Among them, the G1/G2/G3 theories and their variants, and the CBS-Q family of methods by Petersson
and coworkers are worth mentioning in particular. In addition to these wavefunction-based approaches, density
functional methods -- aside from their great popularity as a general tool
for practical computational chemistry -- have gained some currency for computational
thermochemistry in the medium accuracy range, as have group equivalent-based models. 
For very large systems, semiempirical methods remain popular.

\pd At the other extreme in terms of system size and accuracy stand brute-force
approaches such as those based on wavefunctions with explicit interelectronic
distances.

\pd Methods such as G3 and CBS-QB3 do reach the goal of "chemical accuracy" 
(generally defined as $\pm$1 kcal/mol) on average, but worst-case errors for problematic
molecules may exceed this criterion by almost an order of magnitude.
In addition, almost all of these approaches involve some level of parame\-terization
and/or empirical correction against experimental data. While this is by and large
possible (albeit not without pitfalls) in the kcal/mol accuracy range
for first-and second-row compounds, experimental data of sub-kcal/mol accuracy
are thin on the ground, and the available data for transition metal compounds are
simply too scarce for this to be a useful approach.

\pd There would thus appear to be room for a more or less "black box" computational
thermochemistry method that has the following properties:
\begin{enumerate}
\item it on average achieves "benchmark accuracy", which we shall arbitrarily
define as one unit of the most common tabulation unit in thermochemical
reference tables, i.e. 1 kJ/mol (0.24 kcal/mol);
\item the worst-case error should not exceed 1 kcal/mol ("chemical accuracy")
except perhaps in intrinsically pathological cases;
\item it is still efficient enough for applications to systems with up to six
heavy atoms on modern workstations;
\item it is entirely devoid of parameters derived from experiment (and hence
from bias towards the systems used for parameterization).
\end{enumerate}
These have been the design goals in our development of the W1 and W2 \\
(Weizmann-1 and Weizmann-2) theories~\cite{w1}.

\pd The usual design philosophy for this type of methods is bottom-up: one starts
with an approximate model, compares results with experiments, analyzes the
deviations, and uses them to determine empirical corrections and/or additional
terms to be added to the model, after which the cycle is repeated if desired.

\pd Our philosophy was instead "top-down". We decomposed the molecular TAE (total
atomization energy: TAE$\rm _e$ at the bottom of the well, TAE$_0$ at absolute zero)
into all components that can reasonably affect it at the kJ/mol level. Then we
carried out exhaustive benchmark calculations on each component {\em separately} for
a representative "training set" of molecules. Finally, for each component
separately, we progressively introduced approximations up to the point where
reproduction of that particular component started deteriorating to an unacceptable
extent. Thus, experimental data entered the picture only at the validation stage, not
at the design stage.

\pd Another philosophical issue centers on whether a method should be a "protocol"
specified down to the last detail (i.e. be truly "black-box"), or whether it should
merely outline a general approach with minor details to be decided on a case-by-case
basis. Obviously a method where empirical parameterization is kept to the absolute
minimum or is absent altogether will offer more `degrees of freedom' in this regard
than the one where a minor change in the protocol would, for consistency, require 
reparameterization against
a large experimental data set. Yet our general guideline was that,
while such choices should be {\em possible} for an experienced computational chemist,
they should not be an  essential part of the process itself.

\vspace{14pt}
\section{\, Steps in the W1 and W2 theories, \\
\null \, \, \hskip -0.5mm and their justification}
\setcounter{equation}{0}
\vspace{8pt}
\pd The more cost-effective W1 theory and the more rigorous W2 theory have
a lot of points in common. Aside from issues relating to the reference
geometry and the zero-point energy, the main difference concerns the
basis sets used in the extrapolation steps for the SCF and the valence 
correlation contribution.

\pd These basis sets belong to the "correlation consistent" family of 
Dunning and coworkers~\cite{Dun89,Dun97ECC}. The correlation consistent (cc)
basis sets, besides being arguably the most compact ones in their accuracy
range~\cite{Martin1992}, have the important property that, by design, they 
treat radial and angular correlation in a balanced way. In addition to the
regular cc-pVnZ (correlation consistent polarized valence n-tuple 
zeta, or VnZ for short) basis sets, several variants have been published. In particular
we note the aug-cc-pVnZ or AVnZ basis sets~\cite{Ken92} for anions (with
the combination of regular cc-pVnZ on hydrogen and aug-cc-pVnZ
on other elements generally being denoted aug$'$-cc-pVnZ~\cite{Del93}, or A$'$VnZ for short),
the MT (Martin-Taylor~\cite{hf,cc}) and cc-pCVnZ~\cite{Dun95} basis sets
for inner-shell correlation, and the cc-pVnZ+1~\cite{sio}, cc-pVnZ+2d1f~\cite{so2}, and 
(most recently) cc-pV(n+d)Z\cite{angela}
basis sets for second-row atoms exhibiting `inner polarization'~\cite{so2} (vide infra).

\pd We consider here the following sequence of correlation consistent basis sets:
A$'$VDZ+2d, A$'$VTZ+2d1f, A$'$VQZ+2d1f, and
A$'$V5Z+2d1f, which we shall denote "small", "medium",
"large", and "extra large" (for first-and second-row compounds,
these basis sets are of $spd$, $spdf$, $spdf$$g$, and $spdf$$gh$ quality,
respectively). W1 theory, then, carries out all extrapolations
using "small", "medium", and "large", while W2 theory employs
"medium", "large", and "extra-large" basis sets.

\pd The W1 and W2 protocols for obtaining the total atomization energy (TAE) of a given
molecule involve the following steps:
\begin{enumerate}
\item Geometry optimization at the B3LYP/VTZ+1 level for W1, and at the CCSD(T)/VQZ+1
level for W2.
\item Extrapolation of the SCF component of TAE from the "small", "medium", and "large"
basis sets (W1) or "medium", "large", and "extra-large" basis sets (W2), by means of
either the geometric extrapolation formula $\rm E(n)=E_\infty+A/B^n$ (old-style) or
the two-point formula $\rm E(n)=E_\infty+A/n^5$ (new-style).
\item Extrapolation of the CCSD valence correlation component of TAE from the
"medium" and "large" basis sets (W1) or from the "large" and "extra-large" basis sets (W2)
employing the two-point formula $\rm E(n)=E_\infty+A/B^\alpha$, where $\alpha=3.22$ (W1)
or 3 exactly (W2).
\item Extrapolation of the contribution to TAE of the
connected triple excitations, (T), from the valence orbitals
using the same formulae as for CCSD, but employing instead the "small" and "medium"
basis sets (W1) or the "medium" and "large" basis sets (W2).
\item The contribution of inner-shell correlation is taken as the difference
between the CCSD(T)/MTsmall TAE with and without constraining the inner-shell orbitals
to be doubly occupied.
\item The scalar relativistic contribution is computed as the first-order 
Darwin and mass-velocity corrections from the ACPF/MTsmall wave function,
including inner-shell correlation.
\item The contribution to TAE of spin-orbit splitting in the constituent atoms
is trivially obtained from a tabulation, while for molecules in degenerate ground
states, CISD/MTsmall spin-orbit splittings are computed (allowing correlation from 
the $2s$ and $2p$ orbitals in second-row atoms).
\item The zero-point vibrational energy ($\rm E_{ZPV}$) is obtained from harmonic B3LYP/VTZ+1 frequencies
scaled by 0.985 in the case of W1 theory. For W2 theory, anharmonic values of $\rm E_{ZPV}$ from quartic
force fields at the CCSD(T)/VQZ+1
(or comparable) level are preferred; where this is not feasible,
the same procedure as for W1 theory is followed as a "fallback solution".
\end{enumerate}
We shall now proceed to explain in detail these steps and the rationale behind them.

\subsection{Reference Geometry}
\vspace{8pt}
\pd Near the equilibrium geometry, dependence of the energy on geometric
displacements is approximately quadratic. As a result, small errors
in the reference geometry will insignificantly affect computed energies,
but more substantial errors (say, several hundredths of an \AA\ in 
covalent bond lengths) will compromise the reliability of a thermochemical
calculation.

\pd For W1 theory, we chose B3LYP~\cite{Bec93,Lee88}
 density functional theory with the VTZ+1 basis set as the level of theory for the reference
geometry, where the +1 suffix denotes the addition to second-row atoms of the highest-exponent
$d$ function from the V5Z basis set~\cite{sio}. For first-row molecules, B3LYP/VTZ bond lengths are generally
within 0.003 \AA\ from experiment~\cite{dft}; for second-row molecules, significant
errors can be seen~\cite{sio,so3} unless a tight $d$ function is added to the basis
set to account for inner polarization (see below). 

\pd For W2 theory, we opted for CCSD(T)/VQZ+1 as the level of theory for reference geometries.
For geometries, the VQZ basis set is known to be close to the one-particle basis
set limit~\cite{basis2,ch}, while the addition of the inner polarization functions
again takes care of inner polarization effects.

\subsection{The SCF Component of TAE}
\vspace{8pt}
\pd For systems devoid of nondynamical correlation effects, this is the
largest individual contribution to the molecular binding energy.
Its basis set convergence is relatively rapid, yet our discussion
will be disproportionately long because a number of the "dramatis
personae" that reappear in the remainder of the story need to be
introduced here.

\pd For the SCF energy, we can -- at least for small systems -- obtain an 
exact answer by means of numerical SCF calculations. There is
substantial empirical evidence that its convergence behavior is exponential.
Jensen studied the SCF convergence behavior of the SCF
energy in H$_2$ \cite{Jensen1} and H$_3^+$ and N$_2$ \cite{Jensen2} and found
clear evidence of geometric convergence behavior in terms of both the maximum 
angular momentum in the basis set and the number of primitives within a given
angular momentum.

\pd Martin and Taylor~\cite{n2h2} compared numerical SCF energies
with extrapolations from calculated SCF/A$'$VQZ, SCF/A$'$V5Z, and SCF/A$'$V6Z
energies using the formula
\begin{equation}
{\rm E(L) = E_\infty + A/B^L}
\end{equation}
(which is equivalent to ${\rm E(L)=E_\infty+A\exp(-BL)}$ originally proposed by
Feller~\cite{Fel92}) and, for a number of number of molecules, found discrepancies of 10 $\rm \mu E_h$ or less between
the numerical and extrapolated values.

\pd Petersson et al. had earlier proposed~\cite{PeterssonSCF} an alternative expression\\
${\rm E(n)=E_\infty +\sum_{l=n+1}^\infty A/(l+1/2)^6}$
in the context of the CBS methods developed in his group. The summation is carried out
numerically in that paper, but in fact an elegant analytical approximation exists for summations of
this type:
\begin{equation}
{\rm \sum_{m=L+1}^\infty A/(m+1/2)^n = \frac{A \psi^{(n-1)}(L+3/2)}{(n-1)!}}\, ,
\end{equation}
where $\rm \psi^{(n)}(x)$ represents the order n polygamma function~\cite{Abr72} of x.
Its asymptotic expansion has the leading terms
\begin{eqnarray}
\rm \psi^{(n)}(x)&=&\rm
(-1)^{n-1}\left[\frac{(n-1)!}{x^n}+\frac{n!}{2x^{n+1}}+O(x^{-n-2})\right]\nonumber\\[12pt]
&=& \rm (-1)^{n-1}\frac{(n-1)!}{(x-1/2)^n}+O(x^{-n-2})\, .
\end{eqnarray}
Hence
\begin{eqnarray}
\rm \frac{A \psi^{(n-1)}(L+3/2)}{(n-1)!}&=&\rm \frac{(-1)^{n-2}A (n-2)!}{(n-1)!
(L+1)^{n-1}}+O(L^{-(n+1)})\nonumber\\[12pt]
&\approx&\rm \frac{A}{(n-1)(L+1)^{n-1}}\, .
\end{eqnarray}
This suggests the simple extrapolation formula 
${\rm E(n)=E_\infty+A/n^5}$, i.e. ${\rm E_\infty=E(n)+\frac{E(n)-E(n-1)}{(n/n-1)^5-1}}$,
where n is identified with the ``n-tuple zetaness'' of the Dunning correlation consistent
VnZ basis sets. (For hydrogen and helium, n equals the maximum angular momentum plus one;
for the main group elements it is equal to the maximum angular momentum). 
While an argumentation in favor of the Petersson-type formula
can be built on the convergence behavior of triplet-coupled pairs, neither this formula nor the geometric one
have a solid formal basis.

\pd Fortunately, convergence on the SCF component of atomization energies 
is even more rapid than for the total energies; Martin and Taylor
found for 14 first-row molecules~\cite{c2h4tae} that differences between unextrapolated 
SCF/A$'$V5Z, geometrical extrapolations from SCF/A$'$V\{T,Q,5\}Z, and $\rm A+B/L^5$ extrapolations
from SCF/A$'$V\{Q,5\}Z results are on the order of 0.01 kcal/mol. For the method that we
designated W2, which uses this basis set sequence, the choice of SCF extrapolation
method is largely a non-issue. For the method that we designated W1, however, the
geometric formula entails the use of results from the comparatively small 
A$'$VDZ basis set, which compromises the reliability of extrapolated SCF limits
in systems with slow basis set convergence. In some cases (see Table 1 in 
Ref. 26), these can lead to errors of several kcal/mol.
In addition, the two-point $\rm A+B/L^5$ formula has the elegant property that it 
becomes immaterial whether the extrapolation is carried out on a reaction
energy or on the individual absolute energies.

\pd In the original W1/W2 paper~\cite{w1}, we opted for the geometric formula in view
of the observed geometric convergence behavior. In a subsequent validation study~\cite{w1w2validate} on
a much wider variety of systems, we however found the two-point formula to be much more
reliable, and we have adopted it henceforth.

\pd Finally, an issue that arises with second-row systems should be addressed. It was first
noted by Bauschlicher and Partridge~\cite{Bau95} that the atomization energy of 
SO$_2$ is exceedingly sensitive to the presence of high-exponent $d$ and $f$ 
functions in the basis set. This phenomenon was ascribed to hypervalence; Martin and
Uzan~\cite{sio}, however, found that the same phenomenon exists in systems that cannot be
considered hypervalent by the wildest stretch of the imagination, like AlF. In addition,
it was found~\cite{so2,so3} that properties other than the energy are affected as well,
with (e.g. in SO$_2$~\cite{so2} and SO$_3$~\cite{so3}) errors of up to
50 cm$^{-1}$ in harmonic frequencies and hundredths of \AA\ in bond lengths unless high-exponent
$d$ and $f$ functions (termed "inner polarization functions" in Ref. 11 
are added to the basis set. 

\pd We should note that inner polarization is strictly an SCF-level effect: while, for instance,
switching from an A$'$VDZ to an A$'$VDZ+2d basis set affects the computed atomization
energy of SO$_3$ by as much as 40 kcal/mol (!), almost all of this effect is seen
in the SCF component of the TAE~\cite{W1prime}. 
In fact, we have recently found~\cite{MartinUnpub} that the effect persists if the 
$(1s,2s,2p)$ orbitals on the second-row atom are all replaced by a pseudopotential.
What is really getting "polarized" here is the inner part of the valence orbitals,
which requires polarizations functions that are much "tighter" (higher-exponent) than 
those required for the outer part of the valence orbital. The fact that these
inner polarization functions are in the same exponent range as the $d$ and $f$ functions
required for correlation out of the $(2s,2p)$ orbitals is merely coincidental; the
"inner polarization" effect has nothing to do with correlation, let alone with
inner-shell correlation.

\pd After extensive numerical experimentation, we have decided~\cite{w1} on the sequence of basis
sets noted above: "small" A$'$VDZ+2d, "medium"
A$'$VTZ+2d1f, "large" A$'$VQZ+2d1f, and "extra large" A$'$V5Z+2d1f.

\pd As the present review was being finalized for publication, we received a preprint
by Dunning et al.~\cite{angela} where new cc-pV(n+d)Z basis sets are proposed for 
the second-row atoms. These basis sets do have just an added tight $d$ function
(hence the acronym) and no tight $f$ functions, 
but the remaining $d$ functions in the underlying cc-pVnZ basis set
are in addition reoptimized.  We are currently investigating
their performance in W1 and W2-type schemes.

\subsection{The CCSD Valence Correlation Component of TAE}
\vspace{8pt}
\pd The valence correlation component of TAE is the only one that can
rival the SCF component in importance. As is well known by now
(and is a logical consequence of the structure of the exact nonrelativistic
Born-Oppenheimer Hamiltonian on one hand, and the use of a Hartree-Fock
reference wavefunction on the other hand),
molecular correlation energies tend to be dominated by double 
excitations and disconnected products thereof. Single excitation
energies become important only in systems with appreciable nondynamical
correlation. Nonetheless, since the number of single-excitation amplitudes is so
small compared to the double-excitation amplitudes, there is no point in
treating them separately.

\pd For all intents and purposes then, we are concerned here with the CCSD
(coupled cluster with all single and double substitutions~\cite{Pur82})
correlation energy. 
Its convergence is excruciatingly slow: Schwartz~\cite{Sch63} showed as early as 1963
that the increments of successive angular momenta $l$ to the second-order 
correlation energy of helium-like atoms converge as 
\begin{equation}
{\rm \Delta E(}l) = {\rm A/(}l+1/2)^4+{\rm B/(}l+1/2)^6+\ldots\label{eq:schwartz}\, .
\end{equation}
His conclusions were generalized to other methods and general pair correlation energies
by Hill~\cite{Hil85} and by Kutzelnigg and Morgan~\cite{Kut92}. 

\pd This clearly spells a rather bleak picture of basis set convergence. Indeed, Martin~\cite{basis2}
showed in 1994 that while convergence of $\sigma$ bond energies appeared in sight at the
CCSD(T)/$s$$p$$d$$f$$g$ level, this did not yet appear to be the case for $\pi$ bond energies.
This earlier study was extended in 1996~\cite{l4} to basis sets of $s$$p$$d$$f$$g$$h$ quality: somewhat
depressingly, residual errors in the binding energies as high as 2 kcal/mol were still found for
small systems.

\pd However, rather than "knuckling under" to Eq.(\ref{eq:schwartz}) at this stage,
we might instead {\em exploit} it for an extrapolation formula. Martin~\cite{l4} suggested
a three-point extrapolation of the form $\rm A+B/(n+1/2)^C$ (where n is identified with
the cardinal number of the cc-pVnZ basis set), and obtained dramatically
improved computed total atomization energies. A slight further improvement
was achieved if the SCF and valence correlation energies -- which have fundamentally
different convergence behaviors -- are extrapolated separately using the respective
appropriate formulae~\cite{c2h4tae}.

\pd The denominator shift of 1/2 was chosen as a compromise between the situation for 
hydrogen and helium (where $\rm n=l+1$ for the cc-pVnZ basis set) 
and main-group elements (where $\rm n=l$). As is
immediately obvious upon series expansion, there is considerable coupling between
the denominator shift and the exponent. As a result, the three-point extrapolation
generally leads to exponents well in excess of three~\cite{l4}. 

\pd Halkier et al.~\cite{Hal98} found that the simple expression $\rm E(L)=E_\infty+A/L^3$
[i.e. $\rm E_\infty=E(L)+\frac{E(L)-E(L-1)}{(L/L-1)^3-1}$] works at least equally well.
In view of its simplicity and the fact that no results with the questionable
A$'$VDZ basis set are required, we have adopted this simple formula for extrapolation 
of the CCSD valence correlation energy in W1 and W2 theories.

\pd For the smaller basis sets used in W1 theory, the regime where the leading $\rm E_\infty+A/L^3$
term dominates convergence behavior 
has not yet been reached, and using the formula in its unmodified
form leads to overestimated (in absolute value) CCSD limits. One unelegant solution
would be the use of three-term extrapolations like $\rm E_\infty+A/L^3+B/L^4$, but in light of the
poor quality of the VDZ basis set this is a most unsatisfactory alternative.
Another alternative is the use of a two-point extrapolation $\rm E_\infty+A/L^\alpha$,
in which $\alpha$ is a fixed empirical parameter. By minimizing the deviation
from the W2 CCSD limit for the so-called W2-1 set of 28 molecules (vide infra), we
determined $\alpha=3.22$, which is the value used in W1 theory and its variants.

\subsection{Connected Triple Excitations: the (T) Valence \\
\null \hskip 1.5mm Correlation Component of TAE}
\vspace{8pt}
\pd It has been well known for some time (e.g.~\cite{Lee95}) that the next component in importance 
is that of connected triple excitations. By far the most cost-effective
way of estimating them has been the quasiperturbative approach known as
CCSD(T) introduced by Raghavachari et al.~\cite{Rag89}, in which the fourth-order and
fifth-order perturbation theory expressions for the most important terms
are used with the converged CCSD amplitudes for the first-order
wavefunction. This account for substantial fractions of the higher-order
contributions; a very recent detailed analysis by Cremer and He~\cite{Cre2001}
suggests that 87, 80, and 72 \%, respectively, of the sixth-, seventh-, and
eighth-order terms appearing in the much more expensive CCSDT-1a method
are included implicitly in CCSD(T).

\pd Nevertheless, the formidable $\rm n^3N^4$ (with n the number of electrons and
N the number of basis functions) cost scaling of the CCSD(T) method 
creates a substantial barrier to applications of methods that require
A$'$V5Z+2d1f basis sets.
However, two things should be kept in mind. First of all, the (T) component
of TAE is a small fraction of the CCSD component, and hence a larger relative
error can be tolerated. Secondly, evidence exists~\cite{Hel97} that basis
set convergence of the (T) contribution is substantially more rapid than
that of the CCSD energy.

\pd As a result, one may justifiably extrapolate the (T) contribution from smaller
basis sets than its CCSD counterpart: in W1 theory, we extrapolate from
the "small" and "medium" basis sets, and in W2 theory from the "medium" and
"large" basis sets. This means that the most extensive basis sets in the
calculations, namely "large" in W1 theory and "extra large" in W2 theory
only require CCSD calculations, which are both much less expensive than
CCSD(T) and much more amenable to direct algorithms such as those described
in Refs. 40-41. 

\subsection{The Inner-Shell Correlation Component of TAE}
\vspace{8pt}
\pd Inner-shell correlation is a substantial part of the absolute correlation energy
even for late first-row systems; for second-row systems, it in 
fact rivals the absolute valence correlation energy in importance. However,
its relative contribution to molecular TAEs is fairly small: in benzene,
for instance, it amounts to less than 0.7 \% of the TAE. Even so, at 7 kcal/mol,
its contribution is important by any reasonable thermochemical standard.
By the same token, a 1 \% relative error in a 7~kcal/mol contribution is
tolerable even by benchmark thermochemistry standards, while the same relative error 
in a 300 kcal/mol contribution would be unacceptable even by the "chemical accuracy"
standards. 

\pd In addition, for thermochemical purposes we are primarily interested in the
core-valence correlation, since we can reasonably expect the core-core
contributions to largely cancel between the molecule and its constituent atoms.
(The partitioning between core-core correlation -- involving excitations only
from inner-shell orbitals -- and core-valence correlation -- involving simultaneous
excitations from valence and inner-shell orbitals -- was first proposed by
Bauschlicher, Langhoff, and Taylor ~\cite{Bau88}).

\pd For these reasons, we feel justified in treating the inner-shell correlation
contribution to TAE as a separate contribution, rather than together with
the valence correlation. There are substantial cost advantages to this: 
rather than having to carry out very elaborate all-electrons-correlated
CCSD(T) calculations
in basis sets near saturation for both valence and inner-shell
correlation, we can limit these costly calculations to a basis set that
is primarily saturated for inner-shell correlation.

\pd Inner-shell correlation contributions for the W2-1 set were studied in some
detail in the original W1/W2 paper, while subsequently, 
Martin, Sundermann, Fast, and Truhlar (MSFT)~\cite{msft} studied inner-shell 
correlation contributions to TAE for 125 molecules spanning the first two rows of the
periodic table. The following 
conclusions can be drawn from these two studies:
(a) the use of the CCSD(T) electron correlation method is absolutely required
for reliable contributions: the use of MP2 or CCSD can lead to underestimates
in the order of 50 \%;
(b) the smallest basis set which gives acceptable agreement with near-basis set
limit contributions is the MTsmall basis set, which is a completely decontracted
cc-pVTZ basis set with $(2d1f)$ additional high-exponent correlation functions;
(c) the effect of including even higher excitations in the correlation treatment
is insignificant.

\pd A tentative explanation for the importance of connected triple excitations 
for the inner-shell contribution to TAE can be found in the need to account
for simultaneously correlating a valence orbital and relaxing an inner-shell
orbital, or conversely, requiring a double and a single excitation simultaneously.

In principle, one could contract at least the few innermost $s$ primitives and 
reduce the basis set further. By leaving the basis set completely uncontracted,
however, we can recycle the integrals and SCF wavefunction for the next step of the 
calculation.

Finally, it is generally advised {\em not} to correlate the very deep-lying $(1s)$
orbitals on second-row elements, as the MTsmall basis set does not have angular
correlation functions in the required exponent range, and in addition the orbitals
concerned are in the same energy range as the $(2s,2p)$ orbitals in third-row main
group elements, for which being able to take a [Ne] core out of the correlation
problem does result in appreciable CPU time savings.

\subsection{Scalar Relativistic Correction}
\vspace{8pt}
\pd The importance of scalar relativistic effects for compounds of transition
metals and/or heavy main group elements is well established by now~\cite{Pyk88}. 
Somewhat surprisingly (at first sight), they may have nontrivial contributions
to the TAE of first-row and second-row systems as well, in particular if several
polar bonds to a group VI or VII element are involved. For instance, in BF$_3$,
SO$_3$, and SiF$_4$,
scalar relativistic effects reduce TAE by 0.7, 1.2, and 1.9 kcal/mol, respectively
-- quantities which clearly matter even if only "chemical accuracy" is
sought. Likewise, in a benchmark study on the electron affinities of the first-and
second-row atoms~\cite{ea} -- where we were able to reproduce the experimental
values to within 0.001 eV on average -- we saw that neglect of the scalar
relativistic contributions increased mean deviation from experiment by more than an
order of magnitude.

\pd Perhaps the simplest and most cost-effective way of treating relativistic
contributions in an all-electron framework is the first-order perturbation theory
of the one-electron Darwin and mass-velocity operators~\cite{Cow76,Mar83}.  
For variational wavefunctions, these contributions can be evaluated very
efficiently as expectation values of one-electron operators.

\pd It has been found repeatedly~\cite{w1,msft,ea} that scalar relativistic contributions
are overestimated by about 20--25 \% in absolute value at the SCF level.
Hence inclusion of electron correlation is essential: we found the ACPF method
(which is both variational and approximately size extensive) to be an excellent
compromise between quality and cost. It is reasonable to suppose that for
a property that becomes more important as one approaches the nucleus, one wants
maximum flexibility of the wavefunction near the nucleus as well as correlation
of all electrons; thus we finally opted for ACPF/MTsmall as our approach of choice.
Typically the cost of the scalar relativistic step is a fairly small fraction
of that of the core correlation step, since only $\rm n^2N^4$ scaling is involved
in the ACPF calculations.

\pd Bauschlicher~\cite{Bau2000} compared a number of approximate approaches for
scalar relativistic effects to Douglas-Kroll quasirelativistic 
CCSD(T) calculations. He found that the ACPF/MTsmall level of theory faithfully reproduces
his more rigorous calculations, while the use of non-size extensive
approaches like CISD leads to serious errors.  For third-row main  group systems, studies by the
same author~\cite{Bau99Ga} indicate that more rigorous approaches may be in order.

\subsection{Spin-Orbit Coupling}
\vspace{8pt}
\pd The other relativistic effect entirely neglected so far is 
the spin-orbit coupling. For systems in nondegenerate states, the only
first-order contribution to TAE comes from the fine structures
in the corresponding atoms. Their effects can trivially be obtained
from the observed electronic spectra, and hence the computational
cost of this correction is fundamentally zero.

\pd For systems in degenerate states, first-order corrections may need
to be computed. In our work~\cite{w1w2validate} we found that this significantly reduced the 
mean absolute error for the G2-1 and G2-2 test sets for ionization 
potentials and electron affinities, in no small part due to the
preponderance of atoms and linear molecules in these sets.
We found that CISD/MTsmall generally yields quite satisfactory
spin-orbit corrections, but that it is advisable to correlate 
the $(2s,2p)$-like electrons in the second-row elements.
For the halogen atoms, convergence of these contributions
with the level of theory was studied in some detail by 
Nicklass et al.~\cite{Nick2000}. These authors came to fundamentally
the same conclusions.

\subsection{The Zero-Point Vibrational Energy}
\vspace{8pt}

\pd It has been noted repeatedly (e.g.~\cite{Gre91,Sco96,Dix2000c6h6}) that one-half the
sum of the harmonic frequencies, $\rm \frac{1}{2}\sum_i{\omega_i\, d_i}$ (with $\rm d_i$
representing the
degeneracy of mode i) generally leads to an overestimate of the $\rm E_{ZPV}$, and
that one-half the sum of the fundamentals, $\rm \frac{1}{2}\sum_i{\nu_i\, d_i}$, generally
leads to an underestimate. In fact, it is easily shown that the average of these
two estimates is a fairly good approximation to the anharmonic $\rm E_{ZPV}$.

\pd For the sake of convenience, we shall restrict ourselves to the case of symmetric tops, 
asymmetric tops being a special case thereof with no degenerate modes.
Including only up to first-order anharmonicities $\rm X_{ij}$, and excluding the small
constant $\rm E_0$, the vibrational energy is given as 
\begin{equation}
\rm G({\bf n},{\bf l}) = \sum_i{\omega_i\, (n_i+\frac{d_i}{2})}+\sum_{i\leq
j}{X_{ij}\, (n_i+\frac{d_i}{2})(n_j+\frac{d_j}{2})}+S({\bf l}) \, ,
\end{equation}
in which S is the splitting term involving the angular momenta $l$ of the
degenerate vibrations, and $\rm n_i$ represents the vibrational quantum number for
mode i. It trivially follows that the zero-point energy ${\rm E_{ZPV}}$ is given by
\begin{equation}
{\rm E_{ZPV} = \sum_i{\omega_i\,\frac{d_i}{2}}+\sum_{i\leq j}{X_{ij}\, \frac{d_i\, d_j}{4}}}\, .
\end{equation}
In addition we find that [introducing the shorthand $\rm G({\bf n},{\bf l})^0\equiv G({\bf n},{\bf l})-G(0)$)]
\begin{eqnarray}
\rm G({\bf n},{\bf l})^0 &=& \rm \sum_i{\omega_i\, n_i}+\sum_{i\leq
j}{X_{ij}\, [(n_i+\frac{d_i}{2})(n_j+\frac{d_j}{2})-\frac{d_id_j}{4}]}+S({\bf l})\nonumber
\\[8pt] &=& \rm \sum_i{\omega_i\, n_i}+\sum_{i\leq
j}{X_{ij}\, [n_in_j+n_i\frac{d_j}{2}+n_j\frac{d_i}{2}]}+S({\bf
l})\nonumber 
\\[8pt] &=& \rm \sum_i{\omega_i\, n_i}+\sum_{i}{X_{ii}\, n_i\, (n_i+d_i)}\nonumber \\[8pt]
&&\quad \quad \qquad \hskip -1.5mm \rm +\, \frac{1}{2}\, \sum_{i\neq j}X_{ij}\, [n_in_j
+n_i\frac{d_j}{2}+n_j\frac{d_i}{2}]+S({\bf
l})\, .
\end{eqnarray}
Now assume only $\rm n_k$ is nonzero, then
\newpage
\begin{eqnarray}
\rm G(n_k,l_k)^0 &=& \rm \omega_k\, n_k+X_{kk}\, n_k\, (n_k+d_k)\nonumber \\[8pt]
&& \qquad \hskip 1.2mm +\rm \frac{1}{2}\sum_{i\neq k}{(X_{ik}+X_{ki})\frac{n_kd_i}{2}}+S(l_k)\nonumber \\[8pt]
&=& \rm \omega_k\, n_k+X_{kk}\, n_k\, (n_k+d_k)+\sum_{i\neq k}{X_{ik}\,
n_k\,\frac{d_i}{2}}+G_{kk}\, l_k^2\, .\nonumber \\
\end{eqnarray}
It then follows that
\begin{eqnarray}
\rm \sum_k{\nu_k\, \frac{d_k}{2}} &=& \rm \sum_k{\omega_k\, \frac{d_k}{2}}+\sum_k{X_{kk}\,
\frac{d_k(1+d_k)}{2}}
+\sum_k\sum_{i\neq k}{X_{ik}\, \frac{d_id_k}{4}}\nonumber \\[8pt]
&& \qquad \qquad \hskip 0.75mm +\rm \sum_{k}^{\rm degen.}{\frac{d_k}{2}\, G_{kk}\, l_k^2}\nonumber
\\[8pt]
&=& \rm \sum_k{\omega_k\, \frac{d_k}{2}}+\sum_k{X_{kk}\, \frac{d_k^2}{2}}+\sum_k{X_{kk}\, \frac{d_k}{2}}
+\sum_{k > i}{X_{ik}\, \frac{d_id_k}{2}}\nonumber\\[8pt]
&& \qquad \qquad \hskip 0.75mm +\rm \sum_{k}^{\rm degen.}{\frac{d_k}{2}\, G_{kk}\, l_k^2}\nonumber
\\[8pt]
&=& \rm \sum_k{\omega_k\, \frac{d_k}{2}}+\sum_{k\geq i}{X_{ik}\,
\frac{d_id_k}{2}}+\sum_k{X_{kk}\, \frac{d_k}{2}}
+\sum_{k}^{\rm degen.}{\frac{d_k}{2}\, G_{kk}\, l_k^2}\, .\nonumber \\
\end{eqnarray}
That is,
\begin{eqnarray}
\rm \sum_k{(\nu_k+\omega_k)\, \frac{d_k}{4}} &=& 
\rm \sum_k{\omega_k\, \frac{d_k}{2}}+\sum_{k \geq i}{X_{ik}\, \frac{d_id_k}{4}}+\sum_k{X_{kk}\, \frac{d_k}{4}}
\nonumber \\ [8pt]
&& \qquad \qquad \hskip 0.75mm +\rm \sum_{k}^{\rm degen.}{\frac{d_k}{4}\, G_{kk}\, l_k^2}\nonumber
\\[8pt] 
&=& \rm {\rm E_{ZPV}} + \sum_k{X_{kk}\, \frac{d_k}{4}} +\sum_{k}^{\rm degen.}{\frac{d_k}{4}\, G_{kk}\, l_k^2}\,
,
\end{eqnarray}
in which the $\rm G_{kk}$ are the diagonal $l$-coupling constants. The last term is generally
negligible. If so desired, the term involving the diagonal anharmonicity constants
can be estimated from anharmonicities in diatomic molecules.

\pd The common practice of scaling computed vibrational frequencies for comparison with 
experimental fundamentals attempts at approximately addressing two issues: (a) the imperfections
of the theoretical model for the harmonic frequency (which for CCSD(T), or even B3LYP, in sufficiently large
basis sets is basically unnecessary); and (b) the anharmonic
contribution to the fundamental. The above analysis suggests that a scaling factor that is intermediate
between those used for reproducing harmonics and fundamentals would be the most appropriate for
anharmonicities.
In the original W1 paper~\cite{w1}, we considered the essentially exact anharmonic
values of $\rm E_{ZPV}$ of the 28 W2-1 molecules (determined from experiment or large basis set CCSD(T)
quartic force field calculations, e.g.~\cite{c2h4} and the references therein) and found the appropriate
scaling
factor for B3LYP/VTZ+1 harmonic frequencies to be 0.985. The largest
individual deviation between the scaled harmonic and exact anharmonic
values of $\rm E_{ZPV}$ was only 0.3 kcal/mol (for PH$_3$).

\pd Some of the above remarks are probably best illustrated by an example. For benzene, a 
B3LYP/TZ2P quartic force field was computed by Handy and coworkers
~\cite{HandyC6H6}. From the published anharmonicity constants (specifically,
the set deperturbed for Fermi resonances closer than 100 cm$^{-1}$), we obtain
an anharmonic $\rm E_{ZPV}$ of 62.04 kcal/mol. For comparison, one-half the sum of
the harmonics comes out 0.9 kcal/mol too high at 62.96 kcal/mol, 
and one-half the sum of the fundamentals comes out 1 kcal/mol too low at 
60.98 kcal/mol. The average of both values, 61.97 kcal/mol, is in excellent
agreement with the anharmonic value, while the W1 estimate accidentally
agrees to within two decimal places with the B3LYP/TZ2P anharmonic value.
From the best available computed harmonic frequencies~\cite{c6h6} and the best 
available experimental fundamentals~\cite{HandyC6H6}, we obtain $\rm E_{ZPV}=62.01$ kcal/mol
or, after correction for the difference between this estimate and the true anharmonic $\rm E_{ZPV}$
at the B3LYP/TZ2P level, 0.07 kcal/mol, we find $\rm E_{ZPV}=62.08$ kcal/mol as possibly the
best estimate. (Note that HF/6-31G* harmonic frequencies scaled by 0.8929, as used in G2 and G3 theories,
yields only 60.33 kcal/mol. In this accuracy range, one certainly cannot indulge in
a 1.7 kcal/mol underestimate in the zero-point energy!)

\pd In a recent benchmark study~\cite{ch2nh} on the CH$_2$=NH molecule, we explicitly computed
a CCSD(T)/VTZ quartic force field at great expense (the low symmetry necessitated
the computation of 2241 energy points in $C_s$ symmetry and 460 additional points in $C_1$ symmetry).
The resulting anharmonic $\rm E_{ZPV}$, 24.69 kcal/mol, is only 0.10 kcal/mol above the scaled
B3LYP/VTZ estimate, 24.59 kcal/mol. At least for fairly rigid molecules, it appears 
hard to justify the additional expense and effort for the anharmonic force field unless it 
were required anyway for other purposes.

\pd If we use B3LYP/VTZ+1 harmonics scaled by 0.985 for the $\rm E_{ZPV}$ rather 
than the actual anharmonic values, mean absolute error at the W1 level
deteriorates from 0.37 to 0.40 kcal/mol, which most users would
regard as insignificant.  At the W2 level, however, we see a somewhat
more noticeable degradation from 0.23 to 0.30 kcal/mol --- if kJ/mol
accuracy is required, literally "every little bit counts". If one is
primarily concerned with keeping the maximum absolute error down,
rather than getting sub-kJ/mol accuracy for individual molecules,
the use of B3LYP/VTZ+1 harmonic values of $\rm E_{ZPV}$ scaled by 0.985 is an
acceptable "fallback solution". The same would appear to be true for
thermochemical properties to which the  $\rm E_{ZPV}$ contribution is smaller than
for the TAE (e.g. ionization potentials, electron affinities, proton
affinities, and the like).

\begin{table}\linespread{1.1}
\caption{\label{tab:tae0}
Comparison of W2 and W1 theories, and their variants for the evaluation 
of TAE$_0$ (kcal/mol) for the W2-1 test set.}
{\small
\begin{tabular}{lrrrrrrrr}
\hline
         &  \multicolumn{2}{c}{Experimental$^a$} & \multicolumn{6}{c}{Deviation (experiment $-$ theory)}   \\
Species  &  \multispan2{\hrulefill}  & \multispan6{\hrulefill}                           \\
         & TAE$_0$  &  $\pm$ (uncert.) &  W2$^b$   &  W2$^c$  & W2h$^d$  & W1 & W1h$\rm ^d$ & W1c \\ 
\hline
 H$_2$       & 103.27  &  0.00\phantom{00}  &  -0.05  &  -0.04  &    &  -0.07  &    &  -0.07\\
 N$_2$       & 225.06  &  0.04\phantom{00}  &  0.36  &  0.45  &    &  0.53  &    &  0.54             \\
 O$_2$       & 117.97  &  0.04\phantom{00}  &  0.64  &  0.68  &    &  0.41  &    &  0.18             \\
 F$_2$       & 36.94  &  0.10\phantom{00}   &  0.60  &  0.78  &    &  0.70  &    &  0.52              \\
 HF          & 135.33  &  0.17\phantom{00}  &  0.02  &  -0.07  &    &  -0.47  &    &  -0.41          \\
 CH          & 79.90  &  0.23\phantom{00}   &  -0.08  &  -0.15  &  -0.14  &  -0.17  &  -0.11  &  -0.37\\
 CO          & 256.16  &  0.12\phantom{00}  &  0.12  &  0.12  &  0.14  &  -0.08  &  -0.06  &  -0.41   \\
 NO          & 149.82  &  0.03\phantom{00}  &  0.47  &  0.54  &    &  0.56  &    &  0.33              \\
 CS          & 169.41  &  0.23\phantom{00}  &  0.30  &  0.31  &  0.32  &  0.77  &  0.95  &  0.46     \\
 SO          & 123.58  &  0.04\phantom{00}  &  -0.02  &  -0.04  &    &  0.52  &    &  0.57           \\
 HCl         & 102.24  &  0.02\phantom{00}  &  -0.04  &  -0.14  &    &  -0.15  &    &  -0.17         \\
 ClF         & 60.36  &  0.01\phantom{00}   &  0.09  &  0.08  &    &  0.15  &    &  0.03             \\
 Cl$_2$      & 57.18  &  0.00\phantom{00}   &  -0.20  &  -0.24  &    &  0.60  &    &  0.50           \\
 HNO         & 196.85  &  0.06\phantom{00}  &  0.38  &  0.37  &    &  0.20  &    &  -0.03            \\
 CO$_2$      & 381.91  &  0.06\phantom{00}  &  0.14  &  0.13  &  0.10  &  -0.37  &  -0.34  &  -0.37\\
 H$_2$O      & 219.35  &  0.12\phantom{00}  &  -0.04  &  -0.14  &    &  -0.55  &    &  -0.58         \\
 H$_2$S      & 173.15  &  0.12\phantom{00}  &  -0.37  &  -0.49  &    &  -0.47  &    &  -0.51         \\
 HOCl        & 156.61  &  0.12\phantom{00}  &  -0.16  &  -0.24  &    &  -0.18  &    &  -0.40        \\
 OCS         & 328.53  &  0.48\phantom{00}  &  -0.19  &  -0.21  &  -0.21  &  -0.01  &  0.11  &  0.10\\
 ClCN        & 279.20  &  0.48\phantom{00}  &  0.41  &  0.52  &  0.78  &  0.78  &  0.91  &  0.82     \\
 SO$_2$      & 253.92  &  0.08\phantom{00}  &  -0.31  &  -0.33  &    &  0.63  &    &  0.81             \\
 CH$_3$      & 289.00  &  0.10\phantom{00}  &  -0.21  &  -0.32  &  -0.38  &  -0.53  &  -0.51  &  -0.39  \\
 NH$_3$      & 276.73  &  0.13\phantom{00}  &  0.13  &  -0.03  &    &  -0.28  &    &  -0.17           \\
 PH$_3$      & 227.13  &  0.41\phantom{00}  &  -0.01  &  0.28  &    &  0.23  &    &  0.05             \\
 C$_2$H$_2$  & 388.90  &  0.24\phantom{00}  &  0.42  &  0.64  &  0.53  &  0.26  &  0.51  &  0.29       \\
 CH$_2$O     & 357.25  &  0.12\phantom{00}  &  -0.27  &  -0.40  &  -0.35  &  -0.59  &  -0.56  &  -0.76\\
 CH$_4$      & 392.51  &  0.14\phantom{00}  &  -0.11  &  -0.13  &  -0.19  &  -0.35  &  -0.47  &  -0.34\\
 C$_2$H$_4$  & 531.91  &  0.17\phantom{00}  &  -0.19  &  -0.31  &  -0.32  &  -0.63  &  -0.41  &  -0.72 \\
\hline
\multicolumn{3}{c}{Mean Absolute Deviation}     &  0.23  &  0.29  &  0.30  &  0.40  &  0.41  &  0.39      \\
\multicolumn{3}{c}{Max. Absolute Deviation}     &  0.64  &  0.78  &  0.78  &  0.78  &  0.95  &  0.82     \\
\hline
\multicolumn{9}{l}{$\rm ^a$ See ~\cite{w1} for experimental references.}\\
\multicolumn{9}{l}{$\rm ^b$ Values of $\rm E_{ZPV}$ derived from anharmonic vibrational frequencies. 
See Ref. 1}\\
\multicolumn{9}{l}{\phantom{$^o$} for details.}\\
\multicolumn{9}{l}{$\rm ^c$ Values of $\rm E_{ZPV}$ derived from B3LYP/VTZ+1 harmonic vibrational 
frequencies }\\ 
\multicolumn{9}{l}{\phantom{$^o$} scaled by 0.985.  Same remark applies to W2h, W1, W1h and W1c
data given.}\\
\multicolumn{9}{l}{$\rm ^d$ For systems where W2h and W1h are equivalent to W2 and W1, respectively,}\\
\multicolumn{9}{l}{\phantom{$^o$} entries have been left blank.}\\
\end{tabular}
}
\end{table}

\vspace{14pt}
\section{\, Performance of W1 and W2 theories}
\setcounter{equation}{0}
\vspace{8pt}
\pd A reliable assessment of the performance of a method in the kJ/mol
accuracy range is, by its very nature, only possible where experimental
data are themselves known to this accuracy. 

\subsection{Atomization Energies (the W2-1 Set)}
\vspace{8pt}
\pd In the original W1/W2 paper~\cite{w1}, we selected a set of 28 first-and
second-row molecules (which we shall call the W2-1 set) containing at most
three nonhydrogen atoms for which (a) the
experimental total atomization energies $\rm \sum D_0$ are available to 
the highest possible accuracy (preferably 0.1 kcal/mol); (b) no strong nondynamical correlation 
effects exist that would hinder the applicability of single-reference
electron correlation methods; (c) near-exact anharmonic values of $\rm E_{ZPV}$ are available
from either experimental anharmonicity constants or highly accurate ab initio
anharmonic force fields.

\pd Results using W1 and W2 theories are shown in Table \ref{tab:tae0}. For W2 theory we find
a mean absolute deviation (MAD) of 0.23 kcal/mol, which further drops to 0.18 kcal/mol
when the NO, O$_2$, and F$_2$ molecules are deleted (all of which have mild
nondynamical correlation in common). Our largest deviation is 0.70 kcal/mol.
We can hence state that W2 meets our design goals.

\pd For W1 theory, MAD is increased to 0.37 kcal/mol (old SCF extrapolation)
or 0.40 kcal/mol (new SCF extrapolation), with the maximum error being 0.78 kcal/mol.
This should be compared with a MAD of 1.25 kcal/mol for G2 theory, 0.89 kcal/mol for G3 theory, 
0.88 kcal/mol for CBS-Q, and 0.61 kcal/mol for CBS-QB3, and the much higher maximum errors of these methods
of 4.90 kcal/mol (SO$_2$), 3.80 kcal/mol (SO$_2$), 3.10 kcal/mol (OCS), and 1.90 kcal/mol (OCS),
respectively. 
While we would prefer to use W2 theory for no-nonsense benchmarking if at
all possible, W1 theory still seems to offer great advantages over the 
other techniques.

\subsection{Electron Affinities (the G2/97 Set)}
\vspace{8pt}
\pd Some representative results can be found in Table \ref{tab:ipea}.
For the G2-1 set of electron affinities, W1 theory has a mean absolute 
error of 0.016 eV~\cite{w1w2validate}. Not unexpectedly -- given the slow basis set convergence
of electron affinities -- the extra effort invested in W2 theory pays off
with a further reduction of the mean absolute error to 0.012 eV. Accuracy 
appears to be limited principally by imperfections in the CCSD(T) method:
for the atoms B--F and Al--Cl, using even larger basis sets we achieve 0.009 eV
at the CCSD(T) level, which decreases to 0.001 eV if approximate full CI 
energies are used. 

\pd Normally W1 theory does not involve diffuse functions on H, Li, Na, Be, and Mg;
not surprisingly, this leads to very poor electron affinities for Li and Na.
Upon switching to W1aug (i.e. using augmented basis sets on all elements), perfect
agreement with experiment is obtained. Within the G2-2 set, substantial discrepancies
between W1 theory and experiment are found for O$_3$ and CH$_2$NC, both of which
are systems with pronounced multireference character. (The same remark applies to a lesser extent 
to FO.)
Scalar relativistic effects almost invariably decrease the electron affinity.
Neglect of spin-orbit splitting leads to significant deterioration in MAD.

\subsection{Ionization Potentials (the G2/97 Set)}
\vspace{8pt}
\pd Some representative results can again be found in Table \ref{tab:ipea}.
At the W1 level, the G2-1 ionization potentials are reproduced with a MAD of only
0.013 eV~\cite{w1w2validate}. No further improvement is seen at the W2 level for this property. 
Note that if the B3LYP/VTZ geometry for CH$_4^+$ is employed, a serious error
is seen for IP(CH$_4$) which disappears when a CCSD(T)/VTZ reference geometry
is used instead. (Only BH \& HLYP
\newpage
\begin{table}\linespread{1.1}
\caption{\label{tab:ipea}
Comparison of W2 and W1 theories, and their variants for the evaluation of electron affinity 
and ionization potential (eV) for selected species from G2-1 test set.}
\begin{tabular}{lrrrrrr}
\hline
         &  \multicolumn{2}{c}{Experimental$^a$} & \multicolumn{4}{c}{Deviation (experiment $-$ theory)}   \\
Species  &  \multispan2{\hrulefill}  & \multispan4{\hrulefill}                           \\
         & Value  &  $\pm$ (uncert.) &  W2  &   W2h  & W1 & W1h  \\
\hline
\multicolumn{7}{c}{Electron Affinities}\\
\hline
  C  &  1.2629  &  0.0003  &  0.007  &  0.041  &  0.011  &  0.210       \\
  Si  &  1.38946  &  0.00006  &  0.010  &  0.081  &  0.011  &  0.060   \\
  CH  &  1.238  &  0.0078  &  0.029  &  0.060  &  0.032  &  0.248      \\
  CH$_2$  &  0.652  &  0.006  &  0.002  &  0.042  &  0.011  &  0.236    \\
  CH$_3$  &  0.08  &  0.03  &  0.034  &  0.088  &  0.051  &  0.284     \\
  SiH  &  1.2771  &  0.0087  &  0.031  &  0.094  &  0.034  &  0.084   \\
  SiH$_2$  &  1.123  &  0.022  &  0.039  &  0.088  &  0.043  &  0.087  \\
  SiH$_3$  &  1.406  &  0.014  &  0.011  &  0.033  &  0.019  &  0.044  \\
  CN  &  3.862  &  0.005  &  -0.026  &  -0.036  &  -0.031  &  -0.023   \\
\hline
\multicolumn{7}{c}{Ionization Potentials}\\
\hline
B  &  8.29802  &  0.00002  &  0.007  &  0.009  &  0.019  &  0.020   \\
C  &  11.2603  &  0.0001  &  0.010  &  -0.002  &  0.012  &  0.012   \\
Al  &  5.986  &  0.001  &  0.023  &  0.022  &  0.024  &  0.025      \\
Si  &  8.15166  &  0.00003  &  0.018  &  -0.004  &  0.021  &  0.022    \\
CH$_4$ (b)  &  12.61  &  0.01  &  -0.033  &  -0.035  &  -0.032  &  -0.035  \\
SiH$_4$  &  11  &  0.02  &  0.006  &  0.006  &  -0.005  &  -0.005  \\
C$_2$H$_2$  &  11.403  &  0.0003  &  -0.004  &  -0.004  &  -0.001  &  0.005  \\
C$_2$H$_4$  &  10.5138  &  0.0006  &  -0.001  &  0.001  &  -0.005  &  0.000  \\
CO  &  14.0142  &  0.0003  &  -0.014  &  -0.013  &  -0.009  &  -0.008   \\
CS  &  11.33  &  0.01  &  -0.017  &  -0.018  &  -0.017  &  -0.016  \\
\hline
\multicolumn{7}{l}{\small $\rm ^a$ See Ref. 26 for experimental references.}\\
\multicolumn{7}{l}{\small $\rm ^b$ CCSD(T)/VTZ geometry. B3LYP/VTZ optimization erroneously yields $D_{2d}$}\\
\multicolumn{7}{l}{\small \phantom{$^o$} structure for cation rather than correct $C_{2v}$
symmetry. See Ref. 26 for details.}\\
\end{tabular}
\end{table}
\vspace{3cm}
\newpage
\noindent \cite{bhlyp} and mPW1K~\cite{Tru2K} correctly predict a $C_{2v}$ structure
for CH$_4^+$; other exchange-correlation functionals wrongly lead to a $D_2$ structure).

\pd Inner-shell correlation contributions are found to be somewhat more important for 
ionization potentials than
for electron affinities, which is understandable in terms of the creation of a
valence `hole' by ionization into which inner-shell electrons can be excited.
Again, inclusion of spin-orbit splitting is worthwhile.

\subsection{Heats of Formation (the G2/97 Set)}
\vspace{8pt}
\pd A detailed discussion and a table can be found in Ref. 26. 
First of all, we note that the mean uncertainty for the experimental values 
in the G2-1 set is itself 0.6 kcal/mol. MAD values for W1 and W2 theory stand 
at 0.6 and 0.5 kcal/mol, respectively, suggesting that these theoretical
methods have a reliability comparable to the experimental data themselves.

\pd For a subset of 27 G2-2 molecules with fairly small experimental uncertainties,
W1 theory had MAD of 0.7 kcal/mol, compared to the average experimental
uncertainty of 0.4 kcal/mol. Some systems exhibit deviations from experiment in excess of
1 kcal/mol: in the cases of BF$_3$ and CF$_4$, very slow basis set convergence
is responsible, and W2 calculations in fact remove nearly all remaining 
disagreement with experiment for the latter system. (The best available value for BF$_3$
is itself a theoretical one, so a comparison would involve circular reasoning.)
Other molecules (NO$_2$ and ClNO) suffer from severe multireference effects.

\subsection{Proton Affinities}
\vspace{8pt}
\pd For proton affinities, W1 theory can basically be considered converged~\cite{w1w2validate}.
The W2 computed values are barely different from their W1 counterparts, and
the latter's MAD of 0.43 kcal/mol is well below the about 1 kcal/mol
uncertainty in the experimental values. W1 theory would appear to be the
tool of choice for the generation of benchmark proton affinity data for
calibration of more approximate approaches.

\vspace{14pt}
\section{\, Variants and simplifications}
\setcounter{equation}{0}
\vspace{8pt}
\subsection{W1$'$ Theory}
\vspace{8pt}
\pd It was noted that the original W1 theory (old-style SCF extrapolation) performed 
considerably more poorly for second-row than for first-row species. This was
ascribed to the lack of balance in the basis sets for second-row atoms used in the
SCF and valence correlation steps of W1; in particular, the A$'$VTZ+2d1f
basis set contains as many "tight" $d$ and $f$ functions as regular ones, which
would appear to be a bit top-heavy. It was proposed to replace the A$'$VTZ+2d1f basis
set by A$'$VTZ+2d, a conclusion borne out by calculations on the SO$_3$ molecule~\cite{W1prime},
which suffers from extreme inner polarization effects and as such provides a good "proving ground".

\pd Compared to its prototype, the modification (the so-called W1$'$ theory) did appear to yield improved
results for second-row molecules. However, in the W1/W2 validation study~\cite{w1w2validate}
we found this to be an artifact of the exaggerated sensitivity of the (old-style)
3-point geometric SCF extrapolation. Use of the new-style $\rm E_\infty+A/L^5$ extrapolation
largely eliminates both the problem and the difference between W1 and W1$'$ theory.

\subsection{W1h and W2h Theories}
\vspace{8pt}
\pd While the need for diffuse-function augmented basis sets for highly
electronegative elements is well established (e.g.~\cite{l4}), it could
be argued that they are not really required on group III and IV elements.
For organic-type molecules in particular, this would result in significant
savings.

\pd We define here W1h and W2h theories, respectively, as the modifications of W1 theory 
for which AVnZ basis sets are only used on elements of groups V, VI, VII, and VIII, but regular 
VnZ basis sets on groups I, II, III, and IV. (The "h" stands for "heteroatom",
as we originally investigated this for organic molecules.) For the purpose of the present
paper, we have repeated the validation calculations described in the previous
section for W1h and W2h theories. (For about half of the systems, W1 and W1h are
trivially equivalent.) Some representative results can be found in Table \ref{tab:tae0}
for atomization energies/heats of formation, and in Table \ref{tab:ipea} for ionization
potentials and electron affinities.

\pd For the heats of formation in the G2-1 set, the largest difference between W1 and W1h
theory is 0.3 kcal/mol for Si$_2$; the average difference is less than 0.1 kcal/mol.
For some of the systems in the G2-2 set, however, differences are more pronounced,
e.g. 0.6 kcal/mol for CF$_4$ and 0.8 kcal/mol for benzene.
(Note that the benzene calculation reported as an example application in 
the original W1 paper~\cite{w1} is in fact a W1h calculation: the remaining small difference between
that reference and the present work is due to the different SCF extrapolations used.)
For the G2-1 heats of formation, W2h and W2 are essentially indistinguishable in quality,
as could reasonably be expected. 

\pd For the G2-1 ionization potentials, the largest differences are 0.005 and 0.006 eV, respectively,
for ethylene and acetylene. Differences in the G2-2 set are likewise small, although 
Si$_2$H$_2$ (0.009 eV) and CH$_3$OF (0.024 eV) stand out. Clearly W1h is of a quality 
comparable to W1 for ionization potentials, and we recommend it as a moderately inexpensive
high-accuracy method for this property. (As noted before, W2 does not represent an improvement
over W1 for ionization potentials, and the same goes for W2h theory.)

\pd For electron affinities, the differences between W1h and W1 are very pronounced, and become
(as expected) particularly large (e.g. 0.284 eV in CH$_3$) for species where none of the atoms 
carry diffuse functions in W1h theory. The differences between W2 and W2h theory are still
quite sizable, and in fact agreement with experiment for W2h is inferior to that for the
less expensive W1 method. In summary, we do not recommend W1h or W2h for electron affinities.

\subsection{A Bond-Equivalent Model for Inner-Shell Correlation}
\vspace{8pt}
\pd  In a pilot W1h calculation on benzene~\cite{w1}, it was found that 85 \% of the 
CPU time was spent on the inner-shell correlation step. Given that this contribution
is about 0.5 \% of the TAE of benzene, the CPU time proportion appears to be lopsided to say
the least.  On the other hand, a contribution of 7 kcal/mol clearly cannot be neglected by any
reasonable standard.
However, inner-shell correlation is by its very nature a much more local phenomenon
than valence correlation, and a relative error of a few percent in such a small contribution
is more tolerable than a corresponding error in the major contributions, Martin, Sundermann, Fast and
Truhlar (MSFT)~\cite{msft} investigated the
applicability of a bond equivalent model.

\begin{table}\linespread{1.1}
\begin{center}
\caption{\label{tab:core}
Comparison of core correlation contributions to TAE$_0$ (kcal/mol) $\, \, \, \, \, \, \, \, \, \,
\, \, \, \, \, \, \, $ for the W2-1 test set.}
\begin{tabular}{lrrrrr}
\hline
Species  &  CCSD(T)/ & CCSD(T)/ & MSFT      & CPP\phantom{0} & CPP\phantom{0} \\
         &   very large$^a$      & MTsmall\phantom{0} & model     & n = 1$^b$ & n = 2$^b$ \\
\hline
H$_2$       &  0.00\phantom{000}  &  0.00\phantom{000}  &  0.00\phantom{0}  &    &                   \\
N$_2$       &  0.75\phantom{000}  &  0.82\phantom{000}  &  0.80\phantom{0}  &  0.74\phantom{0}  &
1.08\phantom{0} \\
O$_2$       &  0.24\phantom{000}  &  0.24\phantom{000}  &  0.50\phantom{0}  &  0.28\phantom{0}  &
0.43\phantom{0} \\
F$_2$       &  -0.09\phantom{000} & -0.08\phantom{000}  &  0.18\phantom{0}  &  0.05\phantom{0}  &
0.06\phantom{0} \\
HF          &  0.18\phantom{000}  &  0.18\phantom{000}  &  0.09\phantom{0}  &  0.10\phantom{0}  &
0.19\phantom{0} \\
CH          &  0.14\phantom{000}  &  0.14\phantom{000}  &  0.30\phantom{0}  &  0.29\phantom{0}  &
0.48\phantom{0} \\
CO          &  0.94\phantom{000}  &  0.90\phantom{000}  &  1.26\phantom{0}  &  0.76\phantom{0}  &
1.12\phantom{0} \\
NO          &  0.40\phantom{000}  &  0.41\phantom{000}  &  0.51\phantom{0}  &  0.46\phantom{0}  &
0.69\phantom{0} \\
CS          &  0.75\phantom{000}  &  0.66\phantom{000}  &  1.08\phantom{0}  &    &                   \\
SO          &  0.46\phantom{000}  &  0.42\phantom{000} &  0.38\phantom{0}  &    &                   \\
HCl         &  0.20\phantom{000}  &  0.15\phantom{000}  &  0.15\phantom{0}  &    &                   \\
ClF         &  0.08\phantom{000}  &  0.09\phantom{000}  &  0.23\phantom{0}  &    &                   \\
Cl$_2$      &  0.19\phantom{000}  &  0.18\phantom{000}  &  0.29\phantom{0}  &    &                   \\
HNO         &  0.40\phantom{000}  &  0.41\phantom{000}  &  0.68\phantom{0}  &  0.41\phantom{0}  &
0.69\phantom{0} \\
CO$_2$      &  1.64\phantom{000}  &  1.67\phantom{000}  &  1.68\phantom{0}  &  1.12\phantom{0}  &
1.88\phantom{0} \\
H$_2$O      &  0.37\phantom{000}  &  0.37\phantom{000}  &  0.36\phantom{0}  &  0.20\phantom{0}  &
0.39\phantom{0} \\
H$_2$S      &  0.34\phantom{000}  &  0.25\phantom{000}  &  0.24\phantom{0}  &    &                   \\
HOCl        &  0.31\phantom{000}  &  0.29\phantom{000}  &  0.50\phantom{0}  &    &                   \\
OCS         &  1.68\phantom{000}  &  1.58\phantom{000}  &  1.49\phantom{0}  &    &                   \\
ClCN        &  1.76\phantom{000}  &  1.71\phantom{000}  &  1.73\phantom{0}  &    &                   \\
SO$_2$      &  0.67\phantom{000}  &  0.78\phantom{000}  &  0.68\phantom{0}  &    &                   \\
CH$_3$      &  1.04\phantom{000}  &  1.04\phantom{000}  &  0.89\phantom{0}  &  0.37\phantom{0}  &
0.84\phantom{0} \\
NH$_3$      &  0.62\phantom{000}  &  0.64\phantom{000}  &  0.49\phantom{0}  &  0.29\phantom{0}  &
0.62\phantom{0} \\
PH$_3$      &  0.30\phantom{000}  &  0.22\phantom{000}  &  0.35\phantom{0}  &    &                   \\
C$_2$H$_2$  &  2.44\phantom{000}  &  2.34\phantom{000}  &  2.38\phantom{0}  &  1.17\phantom{0}  &
2.17\phantom{0} \\
CH$_2$O     &  1.25\phantom{000}  &  1.26\phantom{000}  &  1.44\phantom{0}  &  0.65\phantom{0}  &
1.24\phantom{0} \\
CH$_4$      &  1.21\phantom{000}  &  1.21\phantom{000}  &  1.19\phantom{0}  &  0.48\phantom{0}  &
1.01\phantom{0} \\
C$_2$H$_4$  &  2.36\phantom{000}  &  2.27\phantom{000}  &  2.38\phantom{0}  &  1.02\phantom{0}  &
2.02\phantom{0}
\\
\hline
\multicolumn{2}{c}{Mean Absolute Deviation}     &  0.04\phantom{000}  &  0.12\phantom{0}  
&  0.39\phantom{0}  &  0.19\phantom{0}  \\
\multicolumn{2}{c}{Max. Absolute Deviation}     &  0.11\phantom{000}  &  0.33\phantom{0}  
&  1.34\phantom{0}  &  0.34\phantom{0}  \\
\hline
C$_6$H$_6$  &    &  7.09\phantom{000}  & 7.13\phantom{0}   &    &  6.30\phantom{0} 
\\
\hline
\multicolumn{6}{l}{\small $\rm^a$ See Ref. 1 for details.}\\
\multicolumn{6}{l}{\small $\rm^b$ See Ref. 60 for details.}\\
\end{tabular}
\end{center}
\end{table}

\pd We started by generating a data base of inner-shell correlation contributions for some
130 molecules that cover the first two rows of the periodic table. In order to reduce
the number of parameters in the model to be fitted, we introduced a Mulliken-type 
approximation for the parameters $\rm D_{AB}\approx (D_{A}+D_{B})/2$. Furthermore we did
retain different parameters for single and multiple bonds, but assumed $\rm D_{A\equiv B}\approx (3/2)D_{A=B}$.

\pd The model (which requires essentially no CPU time)
was found to work very satisfactorily; its performance for the W2-1 set can be
seen in Table \ref{tab:core}. Somewhat to our surprise, we found that the same model
performs reasonably well when applied to the scalar relativistic contributions,
albeit with larger individual deviations.

\pd It was recently suggested by Nicklass and Peterson~\cite{Nic98} that the
use of core polarization potentials (CPPs)~\cite{cpp} could be an inexpensive
and effective way to account for the effects of inner shell correlation. 
The great potential advantage of this indeed rather inexpensive method over the MSFT
bond-equivalent model is that it does not depend on any explicit connectivity
information. 
The different approximate treatments of inner-shell correlation are compared
with large-scale CCSD(T) results for the W2-1 set in Table \ref{tab:core}. As seen there, while
the CPP approach is indeed quite promising (clearly superior to MP2 calculations,
for instance), it clearly requires further refinement.
The MSFT bond-equivalent model in fact outperforms all other
approximate methods, with a computational cost that is essentially nil.

\subsection{Reduced-Cost Approaches to the Scalar Relativistic \\
\null \hskip 1.75mm Correction}
\vspace{8pt}
\pd The fact that the additivity model for the scalar relativistic correction
worked {\em at all} is a pleasant surprise: yet alternatives clearly merit exploration.
As noted above, the SCF-level scalar relativistic contributions of Kedziora
et al.~\cite{Ked2000} are systematically overestimated. One possibility which suggests
itself then would be applying a scaling factor to the SCF values: we have considered
this approach for the set of 120 molecules for which ACPF/MTsmall data were generated by
MSFT for the purposes of parameterizing their empirical model.
However, rather than following the more elaborate approach of Kedziora et al., we simply 
evaluated the first-order Darwin and mass velocity corrections by perturbation theory.
We considered variation
of the basis set, and found not surprisingly that typical contracted VnZ basis sets
are insufficiently flexible in the core region. We found VTZuc+1 (where VTZuc stands for an
uncontracted cc-pVTZ basis set) to be the best compromise between cost and quality.

\begin{table}\linespread{1.1}
\caption{\label{tab:rel}
Comparison of scalar relativistic effect contributions to TAE$_0$ (kcal/mol) for the W2-1 test
set.}
\begin{tabular}{lrrrr}
\hline
Species  &  ACPF/\phantom{0} & MSFT  & B3LYP/\phantom{0:}   & SCF/\phantom{0O:} \\
         &  MTsmall  & model & VTZuc+1\phantom{0}  & VTZuc+1\phantom{0} \\
         &           &       & scaled 0.896 & scaled 0.788\\
\hline
H$_2$       &      0.00\phantom{0:}  &  0.00    &  0.00\phantom{000}  &   0.00\phantom{000} \\
N$_2$       &     -0.11\phantom{0:}  &  -0.14  & -0.15\phantom{000}  &  -0.16\phantom{000} \\
O$_2$       &     -0.15\phantom{0:}  &  -0.30  & -0.18\phantom{000}  &  -0.22\phantom{000} \\
F$_2$       &      0.03\phantom{0:}  &  -0.37   & -0.04\phantom{000}  &  -0.09\phantom{000} \\
HF          &     -0.20\phantom{0:}  &  -0.19  & -0.18\phantom{000}  &  -0.20\phantom{000} \\
CH          &     -0.03\phantom{0:}  &  -0.05  & -0.04\phantom{000}  &  -0.04\phantom{000} \\
CO          &     -0.14\phantom{0:}  &  -0.33  & -0.17\phantom{000}  &  -0.19\phantom{000} \\
NO          &     -0.16\phantom{0:}  &  -0.20  & -0.20\phantom{000}  &  -0.22\phantom{000} \\
CS          &     -0.15\phantom{0:}  &  -0.29  & -0.21\phantom{000}  &  -0.25\phantom{000} \\
SO          &     -0.31\phantom{0:}  &  -0.27  & -0.34\phantom{000}  &  -0.40\phantom{000} \\
HCl         &     -0.26\phantom{0:}  &  -0.17  & -0.25\phantom{000}  &  -0.26\phantom{000} \\
ClF         &     -0.12\phantom{0:}  &  -0.35  & -0.16\phantom{000}  &  -0.23\phantom{000} \\
Cl$_2$      &     -0.15\phantom{0:}  &  -0.34  & -0.19\phantom{000}  &  -0.26\phantom{000} \\
HNO         &     -0.24\phantom{0:}  &  -0.28  & -0.27\phantom{000}  &  -0.29\phantom{000} \\
CO$_2$      &     -0.45\phantom{0:}  &  -0.44  & -0.48\phantom{000}  &  -0.50\phantom{000} \\
H$_2$O      &     -0.26\phantom{0:}  &  -0.26  & -0.25\phantom{000}  &  -0.26\phantom{000} \\
H$_2$S      &     -0.41\phantom{0:}  &  -0.43  & -0.39\phantom{000}  &  -0.40\phantom{000} \\
HOCl        &     -0.28\phantom{0:}  &  -0.43  & -0.31\phantom{000}  &  -0.37\phantom{000} \\
OCS         &     -0.53\phantom{0:}  &  -0.41  & -0.57\phantom{000}  &  -0.57\phantom{000} \\
ClCN        &     -0.43\phantom{0:}  &  -0.40  & -0.47\phantom{000}  &  -0.47\phantom{000} \\
SO$_2$      &     -0.71\phantom{0:}  &  -0.61  & -0.79\phantom{000}  &  -0.90\phantom{000} \\
CH$_3$      &     -0.17\phantom{0:}  &  -0.14  & -0.17\phantom{000}  &  -0.16\phantom{000} \\
NH$_3$      &     -0.25\phantom{0:}  &  -0.24  & -0.25\phantom{000}  &  -0.24\phantom{000} \\
PH$_3$      &     -0.46\phantom{0:}  &  -0.60  & -0.45\phantom{000}  &  -0.46\phantom{000} \\
C$_2$H$_2$  &     -0.27\phantom{0:}  &  -0.31  & -0.28\phantom{000}  &  -0.26\phantom{000} \\
CH$_2$O     &     -0.32\phantom{0:}  &  -0.32  & -0.33\phantom{000}  &  -0.34\phantom{000} \\
CH$_4$      &     -0.19\phantom{0:}  &  -0.19  & -0.19\phantom{000}  &  -0.18\phantom{000} \\
C$_2$H$_4$  &     -0.33\phantom{0:}  &  -0.34  & -0.33\phantom{000}  &  -0.31\phantom{000} \\
\hline
\multicolumn{2}{c}{Mean Absolute Deviation}      &  0.08  &  0.03\phantom{000}  &  0.05\phantom{000}      \\
\multicolumn{2}{c}{Max. Absolute Deviation}      &  0.40  &  0.08\phantom{000}  &  0.20\phantom{000}     \\
\hline
\end{tabular}

\end{table}

\pd The best scale factor in the least-squares
sense is 0.788; while the mean absolute error of 0.04 kcal/mol is more than acceptable,
the maximum absolute error of 0.20 kcal/mol (for SO$_2$) is somewhat disappointing.
Representative results (for the W2-1 set) can be found in Table \ref{tab:rel}.

\pd This error can be considerably reduced, at very little cost, by employing B3LYP
density functional theory instead of SCF. The scale factor, 0.896, is much closer to
unity, and both mean and maximum absolute errors are cut in half compared to
the scaled SCF level corrections. (The largest errors in the 120-molecule data
set are 0.10  kcal/mol for P$_2$ and 0.09 kcal/mol for BeO.) It could in fact
be argued that the remaining discrepancy between the scaled B3LYP/cc-pVTZuc+1
values is on the same order of magnitude as the uncertainty in the ACPF/MTsmall
values themselves.

\subsection{W1c Theory}
\vspace{8pt}
\pd Here we propose a new reduced-cost variant of W1 theory which we shall denote
W1c (for "cheap"), with W1ch theory being derived analogously from W1h theory.
Specifically, the core correlation and scalar relativistic steps are replaced by the
approximations outlined in the previous two sections, i.e. the MSFT bond additivity
model for inner-shell correlation and scaled B3LYP/cc-pVTZuc+1 Darwin and mass-velocity
corrections. Representative results (for the W2-1 set) can be seen in Table \ref{tab:tae0}; complete
data for the molecules in the G2-1 and G2-2 sets are available through the World Wide Web
as supplementary material~\cite{suppmat} to the present paper. 

\pd As seen in Table \ref{tab:tae0}, W1c is an acceptable "fallback solution" for systems for which
W1 calculations are not feasible because of the number of inner-shell orbitals;
for heats of formation and certainly for ionization potentials, W1ch offers a significant
further cost reduction over W1h at a negligible loss in accuracy.

\subsection{Detecting Problems}
\vspace{8pt}
\pd While CCSD and especially CCSD(T) are known~\cite{Lee95} to be
less sensitive to nondynamical correlation
effects than low-order perturbation theoretical methods, some sensitivity 
remains, and deterioration of W1 and W2 results is to be expected for
systems that exhibit severe nondynamical correlation character. A number
of indicators exist for this, such as the ${\cal T}_1$ diagnostic of Lee
and Taylor~\cite{Lee89}, the size of the largest amplitudes in the converged
CCSD wavefunction, and natural orbital occupations of the frontier orbitals.

\pd One pragmatic criterion which we have found to be very useful is the 
percentage of the TAE that gets recovered at the SCF level. For systems
that are wholly dominated by dynamical correlation, like CH$_4$ and
H$_2$, this proportion exceeds 80 \%, while it drops to 50 \% for the N$_2$
molecule, 
O$_2$ is only barely bound at the SCF level, and F$_2$ is
even metastable. In the W1/W2 validation paper~\cite{w1w2validate},
we invariably found that large deviations from what appeared to be reliable 
experimental data tend to be associated with strong nondynamical correlation, and
a small SCF component of TAE (e.g. 27 \% for NO$_2$, \-32 \% for F$_2$O, and  15 \% for  ClO).

\pd Would the use of full CCSDT~\cite{ccsdt} energies, instead of their quasi\-per\-tur\-ba\-tive-triples
CCSD(T) counterparts,  solve the problem?
Our experience has taught us that this 
generally leads to a {\em deterioration} of the 
results; it has been shown (e.g.~\cite{Bak2000}) that the excellent performance of CCSD(T) for
binding energies is at least in part due to error compensation between
partial neglect of higher-order $\rm T_3$ effects and complete neglect of $\rm T_4$ effects.
Unfortunately, explicit treatment of $\rm T_4$ (connected quadruple excitations) is 
at present not feasible for practical-sized systems.

\pd For some very small systems (e.g. Be$_2$~\cite{be2} and
OH/OH$^-$~\cite{oh-}), we have considered what one might term W1CAS and W2CAS, in which the
CCSD(T) calculations were replaced by full valence (or larger) CAS-ACPF calculations.
The SCF extrapolation was then applied to the CASSCF (i.e. Hartree-Fock 
plus
static correlation) energy, and the CCSD/CCSD(T) extrapolation to the dynamical
correlation energy only. Aside from limited applicability due to the explosive 
increase in the number of reference configurations with the number of atoms, 
the formal objection of course applies that any separation between "internal"
and "external" orbital spaces is to a large extent arbitrary.

\pd Common sense also suggests that the larger the "gap" being bridged by the extrapolation
from the actual computed number with the largest basis set to the hypothetical basis
set limit, the larger the uncertainty in the latter will be. (See the example of
benzene in section 5.3.)

\pd Finally, the GIGO ("garbage in, garbage out") theorem applies here as well as in
any other matter. For instance, if a B3LYP/cc-pVTZ+1 reference geometry is used for a
system where the B3LYP geometry is known to be qualitatively wrong (such as CH$_4^+$),
the computed W1 energetics will not be very reliable either.
\vspace{14pt}
\section{\, Example applications}
\setcounter{equation}{0}
\vspace{8pt}
\subsection{Heats of Vaporization of Boron and Silicon}
\vspace{8pt}
\pd First-principle computation of gas-phase molecular heats of formation
by definition requires the gas-phase heats of formation of
the elements:
\begin{eqnarray}
\rm \Delta H^\circ_{f,T}(\hbox{X$_k$Y$_l\cdots$})
   &-& \rm k\,  \Delta H^\circ_{f,T}(\hbox{X})
   - l \, \Delta H^\circ_{f,T}(\hbox{Y}) - \cdots \nonumber\\[3pt]
=  \rm  E_T(\hbox{X$_k$Y$_l\cdots$}) &+& \rm RT\,  (1-k-l-\cdots)
   - k\, E_T(\hbox{X}) - l\, E_T(\hbox{Y}) - \cdots \, \, . \nonumber\\ 
\end{eqnarray}

\pd Somewhat disappointingly, the values of $\rm \Delta H^\circ_f$[A($g$)] of some first- and
second-row elements A (notably boron and silicon) are not precisely known
because of a variety of experimental difficulties.
However, well-established precise heats of formation
of BF$_3$($g$)~\cite{Cod89} and SiF$_4$~\cite{OHare} are available that do not
involve the heats of vaporization of boron and silicon in their determination.
Thus, if accurate computed TAE$_0$ values of BF$_3$ and SiF$_4$ were available,
then, in combination with the established value~\cite{Hub79} of $\rm D_0$(F$_2$),
the quantities sought for could be derived from a thermochemical cycle.
These were obtained by means of W2 theory 
for BF$_3$~\cite{bf3cwb} and for SiF$_4$~\cite{sif4}. The
final recommended values are $\rm \Delta H^\circ_{f,0}$[B($g$)] = 135.1$\pm$0.75 kcal/mol
and $\rm \Delta H^\circ_{f,0}$[Si($g$)] = 107.15$\pm$0.38 kcal/mol. The boron value is
about 2 kcal/mol higher than the CODATA recommended value and in between
a recent evaluation by Hildenbrand~\cite{Hildenbrand} and a 1977 measurement
by Storms and Mueller~\cite{Sto77}.  The silicon value is slightly higher than 
the 
CODATA recommended value, and with a much smaller uncertainty. We note in
passing that one of the first arguments for revision of $\rm \Delta H^\circ_{f,0}$[B($g$)]
and $\rm \Delta H^\circ_{f,0}$[Si($g$)] was given in~\cite{Och95} on computational (CBS-Q) grounds.

\subsection{Validating DFT Methods for Transition States: \\
\null \, the Walden Inversion}
\vspace{8pt}
\pd It is well known (e.g.~\cite{Bak95,And95}) that the prediction of reaction
barrier heights is one of the main "Achilles' heels" of density functional
theory. For instance~\cite{Ada98}, for the prototype S$_N$2 reaction,
\begin{equation}
{\rm X}^- + {\rm CH}_3{\rm Y} \rightarrow {\rm CH}_3{\rm X} + {\rm Y}^-\, ,
\end{equation}
B3LYP predicts a negative overall barrier if $\rm X=Y=Cl$ (i.e. a barrier between
the entry and exit ion-molecule complexes that lies below the entrance
channel). Adamo and Barone~\cite{Ada98} demonstrated that their new mPW1PW91
(modified Perdew-Wang) functional at least yields the correct
sign for this problem.

\pd In Ref. 80 we carried out a W1 and W2 investigation for all six
cases with X,Y$\in$\{F,~Cl,~Br\}, in order to assess the performance 
of a number of DFT exchange-correlation functionals. W2 is in excellent
agreement with experiment where reliable experimental data are available; in
some other cases, the W1 calculations either suggest revisions or provide the only
reliable data available (see Ref. 80 for details).

\pd Of the different exchange-correlation functionals considered, the new
mPW1K~\cite{Tru2K} functional of Truhlar and coworkers appears to yield
the best performance among "hybrid" functionals (i.e. those including a
fraction of exact exchange), followed by BH\&HLYP (a half-and-half mixture~\cite{bhlyp} of
Hartree-Fock and
Becke 1988 exchange~\cite{Bec88} with Lee-Yang-Parr correlation). Among "pure DFT"
functionals, the best performance is delivered by HCTH-120~\cite{hcth120} (the
120-molecule reparameterization of the Hamprecht-Cohen-Tozer-Handy functional).
(We note in passing that this latter functional was parameterized entirely against 
ab initio data.) The G2 data of Pross et al.~\cite{Pro96}, despite some
quantitative discrepancies, is qualitatively in perfect agreement with
W1 theory.

\pd We also note that in one case (F, Br) it was impossible to obtain all required
stationary points at the B3LYP level, since the F{$\cdots$}CH$_3$Br minimum does not
show up at all at this level. Only mPW1K and BH\&HLYP find this stationary point,
as does CCSD(T).
\subsection{Benzene as a "Stress Test" of the Method}
\vspace{8pt}
\pd As an illustrative example of "stress-testing" W1 and W2 theory, we
shall consider the benzene molecule\cite{c6h6w2}. The most accurate calculation
we were able to carry out is at the W2h level: the rate-determining
step was the direct CCSD/cc-pV5Z calculation (30 electrons correlated,
876 basis functions, carried out in the $D_{2h}$ subgroup of $D_{6h}$)
which took nearly two weeks on an Alpha EV67/667 MHz CPU. Relevant
results are collected in Table \ref{tab:c6h6}.

\begin{table}[h]
\caption{Individual components in W1h, W1, and W2h total atomization energy
{\em cum} heat of formation of benzene. All data in kcal/mol.$\rm ^a$\label{tab:c6h6}}
{\footnotesize
\begin{tabular}{lrrrrrr}
\hline
Reference geometry    & \multicolumn{4}{c}{B3LYP/cc-pVTZ} & \multicolumn{2}{r}{CCSD(T)/cc-pVQZ}\\
              & \multicolumn{2}{c}{W1h} & \multicolumn{2}{c}{W1} & \multicolumn{2}{c}{W2h}\\
\hline
SCF           & VDZ & 1024.19 & A$'$VDZ & 1024.59 & \phantom{000:}VTZ & 1042.16 \\
              & VTZ & 1042.10 & A$'$VTZ & 1042.62 & \phantom{000:}VQZ & 1044.62 \\
              & VQZ & 1044.56 & A$'$VQZ & 1044.84 & \phantom{000:}V5Z & 1045.30 \\
old-style    & V$\infty$Z & 1044.95 & V$\infty$Z & 1045.15 & \phantom{000:}V$\infty$Z & 1045.56 \\
new-style    & V$\infty$Z & 1045.33 & V$\infty$Z & 1045.53 & \phantom{000:}V$\infty$Z & 1045.63 \\
CCSD          & VDZ &  225.94 & A$'$VDZ &  226.11 & \phantom{000:}VTZ &  265.49\\
              & VTZ &  265.55 & A$'$VTZ &  268.44 & \phantom{000:}VQZ &  280.91\\
              & VQZ &  280.97 & A$'$VQZ &  282.39 & \phantom{000:}V5Z &  285.72\\
       & V$\infty$Z &  291.08 & V$\infty$Z &  291.53 & \phantom{000:}V$\infty$Z & 290.77 \\
(T)           & VDZ &   18.72 & A$'$VDZ &   19.64 & \phantom{000:}VTZ &   24.41\\
              & VTZ &   24.42 & A$'$VTZ &   24.78 & \phantom{000:}VQZ &   25.74\\
       & V$\infty$Z &   26.55 & V$\infty$Z &   26.69 &\phantom{000:} V$\infty$Z &  26.71 \\
\multicolumn{2}{l}{Inner-shell correlation}  & 7.09 &  & 7.08 &  & 7.10 \\
\multicolumn{2}{l}{Darwin and mass-velocity} & -0.99 &  & -0.99 &  & -0.99 \\
\multicolumn{2}{l}{Spin-orbit coupling} & -0.51 &  & -0.51 &  & -0.51 \\
TAE$_e$ &  & 1368.54 &  & 1369.33 &  & 1368.71\\
$\rm E_{ZPV}$ &  & 62.04 &  & 62.04 &  & 62.04\\
TAE$_0$ &  & 1306.49 &  & 1307.29 &  & 1306.67\\
$\rm \Delta H^\circ_{f, 0 K}$[C$_6$H$_6$(g)] &  & 23.18 &  & 22.39 &  & 23.01\\
$\rm \Delta [H_{298.15}-H_0]$ &  & -4.24 &  & -4.24 &  & -4.24\\
$\rm \Delta H^\circ_{f, 298.15 K}$[C$_6$H$_6$(g)] &  & 18.95 &  & 18.15 &  & 18.78\\
\hline
\multicolumn{7}{l}{$\rm ^a$ Lower level TAE$_0$: 1301.9 (G2), 1305.2 (G3), 1303.7 (CBS-QB3), and 1304.3
(CBS-Q)}\\
\multicolumn{7}{l}{\phantom{$^o$} kcal/mol. Experiment: $\rm \Delta H^\circ_{f, 0 K}$[C$_6$H$_6$(g)] =
24.0$\pm$0.2 kcal/mol. [J. B. Pedley, {\it Thermo-}}\\
\multicolumn{7}{l}{\phantom{$^o$} {\it dynamic Data and Structures of Organic Compounds}
(Thermodynamics Research Cen-}\\
\multicolumn{7}{l}{\phantom{$^o$} ter College Station, TX, 1994); Vol. 1.] $\, $ This 
standard enthalpy of formation produces}\\
\multicolumn{7}{l}{\phantom{$^o$} TAE$_0$ = 1305.7$\pm$0.7 kcal/mol, where the 
uncertainty equals $\sqrt{0.2^2+(6\times0.11)^2}$; 0.11}\\
\multicolumn{7}{l}{\phantom{$^o$} kcal/mol being the uncertainty in the CODATA $\rm \Delta
H^\circ_{f,0}$ [C(g)] = 169.98$\pm$0.11 kcal/mol}\\
\multicolumn{7}{l}{\phantom{$^o$} \cite{Cod89}. (The uncertainty in $\rm \Delta H^\circ_{f,0}$
[H(g)] is negligible.)}\\
\end{tabular}
}
\end{table}

\pd At first sight, the disagreement between the computed W2h value of $\rm \Delta H^\circ_{f, 0
K}$ = 23.0 kcal/mol
and the experimental value of 24.0$\pm$0.2 kcal/mol seems disheartening. 
(Note that it "errs" on the other side as the most recent previous benchmark
calculation~\cite{Dix2000c6h6},
24.7$\pm$0.3 kcal/mol, using similar-sized basis sets as W1 theory.)
However, the comparison with experiment
is not entirely "fair" since it neglects the experimental uncertainties in the atomic heats of
formation
required to convert an atomization energy into a heat of formation (or vice versa). Combining these
with the experimental $\rm \Delta H^\circ_{f, 0 K}$ leads to an experimentally derived 
TAE$_0=1305.7\pm 0.7$ kcal/mol, where the uncertainty is dominated by six times that in the
heat of vaporization of graphite. In other words, our calculated TAE$_0=1306.7$ kcal/mol is
only 0.3 kcal/mol removed from the upper end of the experimental uncertainty interval.
(After all, an error of 0.02 \% seems to be a bit much to ask for.)  

\pd Secondly, let us consider the "gaps" bridged by the extrapolations. For the SCF component,
that gap is a very reasonable 0.3 kcal/mol (0.03 \%), but for the CCSD valence correlation
component this rises to 5 kcal/mol (1.7 \%) while for the connected triple excitations
contribution it amounts to 1 kcal/mol (3.7 \% --- note however that a smaller basis set
is being used than for CCSD). It is clear that the extrapolations are indispensable to
obtain even a {\em useful} result, let alone an accurate one, even with such large basis
sets. 

\pd Inner-shell correlation, at 7 kcal/mol, is of quite nontrivial importance, but even scalar relativistic
effects (at 1 kcal/mol) cannot be ignored. And manifestly, even a 2 \% error in a 62 kcal/mol
zero-point
vibrational energy would be unacceptable. 

\pd Let us now consider the more approximate results. While W1h coincidentally agrees to better than 0.2 
kcal/mol with the W2h result, W1 deviates from the latter by 0.6 kcal/mol. Note, however, that in 
W1h theory, the extrapolations bridge gaps of 0.8 (SCF), 10.1 (CCSD), and 2.1 (T) kcal/mol,
the corresponding amounts for W1 theory being 0.7, 9.1, and 1.9 kcal/mol, respectively. 
Common sense suggests that if extrapolations account for 13.0 (W1h) and 11.7 (W1) kcal/mol, then
a discrepancy of 1 kcal/mol should not come as a surprise --- in fact, the relatively good agreement
between the two sets of numbers and the more rigorous W2h result (total extrapolation: 6.3 kcal/mol)
testifies, if anything, to the robustness of the method.

\pd As for the difference of about 0.4 kcal/mol between the old-style and new-style SCF 
extrapolations in W1h and W1 theories, comparison with the W2h SCF limits clearly suggests
the new-style extrapolation to be the more reliable one. (The two extrapolations yield basically
the same result in W2h.) This should not be seen as an indication that the $\rm E_\infty+A/L^5$ formula
is somehow better founded theoretically, but rather as an example of why reliance on (aug-)cc-pVDZ
data should be avoided if at all possible. Users who prefer the geometric extrapolation for the
SCF component could consider carrying out a direct SCF calculation in the "extra large" (i.e. V5Z) basis set
and applying the $\rm E_\infty+A/B^L$ extrapolation to the "medium", "large", and "extra large" SCF
data.

\vspace{14pt}
\section{\, Conclusions and Prospects}
\vspace{8pt}
\pd W1/W2 theory and their variants would appear to 
represent a valuable addition to the computational chemist's toolbox,
both for applications that require high-accuracy energetics for 
small molecules and as a potential source of parameterization data
for more approximate methods.
The extra cost of W2 theory (compared to W1 theory) does appear to
translate into better results for heats of formation and electron affinities,
but does not appear to be justified for ionization potentials and
proton affinities, for which the W1 approach yields
basically converged results. Explicit calculation of anharmonic
zero-point energies (as opposed to scaling of harmonic ones) does
lead to a further improvement in the quality of W2 heats of formation;
at the W1 level, the improvement is not sufficiently noticeable to justify
the extra expense and difficulty.

\pd Of the various reduced-cost variants introduced in this paper, W2h performs
basically as accurately as to W2 for heats of formation. Likewise, W1h is essentially
as good as W1 theory for ionization potentials, and almost as good for 
heats of formation. Neither method is recommended for electron affinities.

\pd In systems where a large number of inner-shell electrons makes the inner-shell 
correlation (and, to a lesser extent, scalar relativistic) steps in W1 and
W2 theory unfeasible, the use of a bond equivalent model for the inner-shell
correlation and scaled B3LYP/cc-pVTZuc+1 scalar relativistic corrections
offers an alternative under the name of W1c and W1ch  theories.

\pd One plan for the future is the extension to heavier element systems; the first
step in this direction has been made recently with the development of the
SDB-cc-pVnZ valence basis sets~\cite{sdb} (for use with the Stuttgart-Dresden-Bonn
relativistic ECPs~\cite{sdbreview}) for third- and fourth-row main group elements.

\pd Further improvement of accuracy, as well as applicability to systems exhibiting
nondynamical correlation, will almost certainly require some level of treatment
of connected quadruple excitations.\\ 

\vspace{3pt}
\acknowledgements
\vspace{8pt}
\pd JM is the incumbent of the Helen and Milton A. Kimmelman Career 
Development Chair. SP acknowledges a Clore Postdoctoral Fellowship.
This research was supported by the {\em Tashtiyot} (Infrastructures)
program of the Ministry of Science and Technology (Israel). The authors
would like to acknowledge helpful discussions with Prof. Peter R. Taylor
(UCSD), Drs. Charles W. Bauschlicher Jr. and Timothy J. Lee (NASA Ames
Research Center), Drs. David A. Dixon and Thom. H. Dunning Jr. (PNNL),
Prof. Martin Head-Gordon (UC Berkeley), and finally Dr. Michael J. Frisch (Gaussian, Inc.),
who originally suggested to us that we consider B3LYP as a reduced-cost alternative
for the scalar relativistic correction. We thank Mark Iron for editorial assistance
and Dr. Angela K. Wilson for a preprint of Ref. 12. 
\endacknowledgements

\end{document}